\documentclass[preprint,nofootinbib]{revtex4}%
\usepackage{amssymb}
\usepackage{amsfonts}
\usepackage{amsmath}
\usepackage{graphicx}
\usepackage[usenames]{color}%
\usepackage{pgfplots}
\pgfplotsset{compat=1.18}

\providecommand{\U}[1]{\protect\rule{.1in}{.1in}}
\definecolor{blue}{rgb}{0,0,1}

\definecolor{red}{rgb}{1,0,0}

\begin{document}
\title{A cubic theory of gravity with quasitopological-like properties for compactified cosmologies}

\author{Vicente Aguilar$^1$, Fabrizio Canfora$^2$, Vicente García$^1$, Alex Giacomini$^3$, Julio Oliva$^1$}

\affiliation{$^1$Departamento de Física, Universidad de Concepción, Casilla, 160--C, Concepción, Chile}
\affiliation{$^2$Centro de Estudios Científicos (CECs), Casilla 1469, Valdivia, Chile and
Facultad de Ingeniería, Universidad San Sebastián,
sede Valdivia, General Lagos 1163, Valdivia 5110693, Chile}
\affiliation{$^3$Instituto de Ciencias Físicas y Matemáticas, Universidad Austral de Chile, Valdivia, Chile}
\begin{abstract}
We identify a cubic combination of curvature invariants in dimension $D=7$ that when added to the Einstein-Hilbert action, yields a healthy cosmological evolution in four dimensions after compactification on a constant-curvature internal space of dimension three. For generic backgrounds it leads to fourth-order field equations, nevertheless, on a warped product of a spatially flat, homogeneous and isotropic four-dimensional universe with a constant-curvature space of radius $b(t)$, the constraint is of first order in the scale factor $a(t)$ and in $b(t)$, while the remaining equations are of second order. Requiring this property uniquely singles out the cubic theory. We study the solutions describing a four-dimensional de Sitter Universe in direct product with an internal space of stabilized radius and evaluate its linear stability under time-dependent perturbations, showing that for a given sign of the curvature of the internal space, stability of the static solution fixes the sign of the higher curvature coupling. We also argue that linearized scalar perturbations depending on the spatial coordinates obey equations with at most two time derivatives. Finally, we numerically evolve the homogeneous and isotropic system non-linearly and present evidence that for a positive bare cosmological term $\Lambda_0$ and a hyperbolic compact space, the singularity formation present in GR in this ansatz is pushed into the future by increasing the higher-curvature coupling.
\end{abstract}
\maketitle

\section{Introduction}
Higher curvature terms, complementing the Einstein-Hilbert Lagrangian,
naturally arise in the context of String Theory when $\alpha^{\prime}$-corrections emerge in a perturbative expansion around the infinite tension
limit \cite{Zwiebach:1985uq}, \cite{Gross:1986iv}, \cite{Metsaev:1987bc}, \cite{Metsaev:1987zx}, \cite{Bergshoeff:1989de}. The computation of the precise form of these corrections, beyond the
first few orders, becomes a remarkable task, which can be restricted and/or
simplified by performing suitable field redefinitions and by considering the
symmetries that the truncated series must present at each order (see e.g. \cite{Hohm:2015doa} and references therein). Due to
such difficulties, it is still relevant to approach the problem in a bottom-up manner, namely by
proposing suitable higher-curvature corrections to GR, which fulfill some
consistency conditions. Following this approach, families of higher-curvature theories have been identified during the last decades, restricting the
lagrangians with different properties that eventually have led to interesting, well-posed frameworks to
evaluate the effects of higher-curvature terms in strong gravity regimes,
cosmology, and in holographic scenarios when the theories are
formulated in asymptotically AdS spacetimes \cite{Oliva:2010eb}-\cite{Bueno:2016lrh}. For example, in \cite{Oliva:2010eb} a
cubic combination in five dimensions was proposed that, in spite of leading to fourth-order field equations for generic backgrounds, fulfills the Birkhoff's
theorem. It turned out that gravitons propagating on AdS in this theory have the same
propagator as that in GR \cite{Myers:2010ru}, with rescaled cosmological and
Newton's constants. This cubic combination was the first of a family of
higher-curvature lagrangians dubbed quasitopological gravities. A quartic
case was identified in \cite{Dehghani:2011vu}, while a family of quintic theories were built in
\cite{Cisterna:2017umf}. It was then realized that theories with such
properties exist at arbitrary order \cite{Bueno:2019ycr} and a recurrence relation for
their construction was provided which can be solved as in \cite{Moreno:2023rfl}.
These quasitopological theories present special properties in spherically
symmetric scenarios, where the field equation in turn imply that the spacetime
must be static (see also the recent \cite{Murcia:2026pmg}, \cite{Bueno:2026wse}).

A natural question then emerges: is it possible to identify suitable higher-curvature Lagrangians, with special properties in other families of relevant
spacetimes? For the cosmological Friedmann-Lema\^{\i}tre-Robertson-Walker (FLRW) ansatz in four dimensions, this
question has been answered positively. In \cite{Arciniega:2018fxj} it was found
that a cubic combination in the Riemann tensor, when added to the
Einstein-Hilbert action, may lead to a first-order equation for the scale
factor $a\left(  t\right)  $, which in turn can be written as a polynomial
equation on the Hubble function $H\left(  t\right)  =\frac{\dot{a}\left(
t\right)  }{a\left(  t\right)  }$. The cubic combination turned out to be the
trivial dimensional reduction of the cubic, five-dimensional,
quasitopological combination as shown in \cite{Cisterna:2018tgx}, and within this approach also in the latter reference, four-dimensional,
higher-curvature combinations with the same properties in the FLRW spacetime were constructed for the quartic and
quintic case. The original work \cite{Arciniega:2018fxj} led to the realm of Geometric Inflation, where these suitable higher-curvature terms contribute to the inflationary period followed by a radiation-dominated era, making the job of the inflaton easier (\cite{Arciniega:2018tnn},\cite{Arciniega:2019oxa},\cite{Edelstein:2020nhg}). The theories with these properties were constructed at all orders in \cite{Arciniega:2018tnn} (for the three-dimensional analysis and more see \cite{Moreno:2023arp}), and it was shown in \cite{Cisterna:2024ksz} that they may generically lead to the emergence of a de Sitter inflationary phase near the singularities of an arbitrary dynamical function controlling the resummation of an infinite series of higher-curvature combinations. Notice that it was originally realized in \cite{Bueno:2024dgm} that in the context of asymptotically flat black holes such a resummation may lead to regular black holes in pure gravity (see also \cite{Aguayo:2025xfi} and \cite{Hennigar:2025yqm} for the natural extension to asymptotically AdS solutions and their thermodynamics).

It is clear that in order to connect higher-dimensional theories with the
observed four-dimensional physics, we need some mechanism of compactification or projection of the higher-dimensional physics into a four-dimensional brane. Such a mechanism could be associated to a (semi-)classical instability as argued in the seminal work \cite{Witten:1981gj}. For example, in \cite{Canfora:2008iu} it was shown that the cubic,
seven-dimensional Lovelock theory permits a compactification to dimension
four, for a suitable choice of the coupling of the theory, in such a manner
that Einstein equations are consistently induced on the four-dimensional
spacetime, providing an example of compactification of a gravitational, higher-curvature
theory with second-order field equations. If one relaxes this latter
requirement, as it is necessary in the truncated $\alpha^{\prime}$ expansion
of String Theory, one might try to explore the set of unstable higher-dimensional
theories that stabilize when compactified to dimension four.

In this work we present a new cubic combination of curvature invariants that leads to fourth-order field equations on generic spacetimes, but reduces to a first-order Friedmann equation on a cosmological ansatz consisting of a four-dimensional FLRW spacetime times a compact manifold of constant curvature $K=\pm1,0$ and radius $b(t)$. Imposing this requirement singles out a unique cubic combination. The model has the properties described above: on a physically relevant ansatz such as a cosmological one, where the evolution in higher dimensions is unstable, compactification leads to a well-posed cosmological evolution in four dimensions. Moreover, the theory singles out four macroscopic directions even when it is formulated in a higher-dimensional spacetime.

The paper is organized as follows. In Section~\ref{sec:model} we introduce the theory and show that imposing second-order equations on a homogeneous and isotropic compactified ansatz fixes the spacetime dimension to be seven, with the cosmological evolution taking place in the four non-compact directions. In Section~\ref{sec:stab} we show that the size of the three compact directions can be stabilized, by analyzing linearized perturbations that preserve homogeneity and isotropy of the four-dimensional cosmology. Section~\ref{sec:scalar} presents a family of scalar perturbations that break homogeneity and isotropy and, remarkably, still lead to equations with at most two time derivatives around the de~Sitter$_4\times\Sigma_K$ solution. In Section~\ref{sec:nonlin} we analyze singularity formation under the nonlinear evolution of the system, and study how the time of formation depends on the higher-curvature coupling and on the initial conditions. Finally, Section~\ref{sec:concl} contains our conclusions and further comments.

\section{The model whose sensible dynamics requires a $4+3$ split of the dimension}\label{sec:model}
The new cubic theory supplementing the Einstein-Hilbert action, reads%
\begin{align}
\mathcal{L} ={}& R - 2\Lambda_0 + d_8\Bigl(
    R^3
    - \frac{504}{73}\, R_{c}{}^{j}{}_{e}{}^{k}\, R^{cdei}\, R_{djik}
    - \frac{1080}{73}\, R^{cd}\, R_{c}{}^{eij}\, R_{deij}
    + \frac{171}{73}\, R\, R_{cdei}\, R^{cdei} \nonumber\\
  &\qquad\quad
    + \frac{1512}{73}\, R^{cd}\, R^{ei}\, R_{cedi}
    + \frac{1008}{73}\, R_{c}{}^{e}\, R^{cd}\, R_{de}
    - \frac{828}{73}\, R_{cd}\, R^{cd}\, R
  \Bigr)\,, \label{lag}
\end{align}
where $d_8$ stands for the higher-curvature coupling with dimensions of $\text{length}^4$. Unless otherwise mentioned, we will deal with finite values of the higher curvature couplings, nevertheless, in the context of string theory $\mathcal{R}^3$ terms emerge in the effective action at the order $\alpha'^2$. Even though cubic terms appear naturally in the bosonic string, in the type II theories they can be brought in by introducing suitable field redefinitions. The potential origin of our new cubic combination \eqref{lag} from a UV-complete framework is beyond the scope of this work and we will not comment any further on this vein.

We are interested in the physics implied by this theory on a compactified cosmological ansatz, namely
\begin{equation}
ds^2=-N(t)^2\,dt^2+a(t)^2\,d\vec{x}_d^{\,2}+b(t)^2\,d\Sigma_p^2\ ,\label{anz}
\end{equation}
where $d\vec{x}_d^{\,2}$ is the flat spatial section of a $(d+1)$-dimensional FLRW spacetime, and $d\Sigma_p^2$ is the line element of a $p$-dimensional Euclidean space of constant curvature $K$. The total spacetime dimension is $D=1+d+p$. We split the coordinates accordingly: the Greek indices $(\mu,\nu,\ldots)$ run over all coordinates, the middle Latin indices $(i,j,\ldots)$ label the $d$ flat directions, and the initial Latin indices $(a,b,\ldots)$ label the $p$ compact directions. The comoving time coordinate is denoted by $t$.

We have included the lapse $N(t)$, which can in principle be gauged away. However, since this ansatz is compatible with the principle of symmetric criticality for diffeomorphism-invariant theories, the action must be evaluated with a non-constant $N(t)$ for the reduction to be consistent. In summary, evaluating the action \eqref{lag} on the ansatz \eqref{anz} and varying the reduced action with respect to the degrees of freedom $(N(t),a(t),b(t))$ consistently yields the correct field equations.

The highest derivative of $a(t)$ appearing in the equation for $N(t)$ is a third derivative with respect to time. In the coefficient of the third derivative, the highest derivative of $b(t)$ is the second. The equation for $N(t)$ then contains a term of the form
\begin{equation}
p\times d\times a^4(t)b^4(t)d_8\left((d-3)b(t)\frac{da(t)}{dt}+(p-3)a(t)\frac{db(t)}{dt}\right)\frac{d^3}{dt^3}a(t)\frac{d^2}{dt^2}b(t)+\ldots=0\ .\label{eqcrucial}
\end{equation}
Requiring the absence of these higher-derivative terms in the constraint, the equation obtained by varying the action with respect to the lapse $N(t)$ immediately imposes $p=3$ and $d=3$, namely that the spacetime dimension must be seven, that there must be three compact dimensions and that the cosmological evolution of the scale factor and the radius of the extended directions is dictated by field equations that do not contain higher derivatives. Indeed, selecting $p=3$ and $d=3$, the constraint and the equations coming from variations with respect to $a(t)$ and $b(t)$, respectively reduce to
\begin{align}\label{constraint}
0=&73\,\frac{\Lambda_0 b^{2}-3K-3\dot b^{2}}{b^{2}}
-219\,H_a^{2}-657\,H_a\frac{\dot b}{b}
-5832\,d_8\,H_a^{3}\,\frac{\dot b\,\big(3K+5\dot b^{2}\big)}{b^{3}},\\[2mm]
\label{del_a} 0=&146\Big(\dot H_a+H_a^{2}+3H_a\frac{\dot b}{b}\Big)+73H_a^{2}
+73\,\frac{3K+3\dot b^{2}+3\,b\,\ddot b-\Lambda_0 b^{2}}{b^{2}}\nonumber\\
&\quad+5832\,d_8\,\frac{H_a}{b^{3}}\Big[2\,\dot b\,\big(K+\dot b^{2}\big)\big(\dot H_a+H_a^{2}\big)
+H_a\big(K+3\dot b^{2}\big)\,\ddot b\Big],\\[2mm]
\label{del_b}0=&219\Big(\dot H_a+2H_a^{2}+2H_a\frac{\dot b}{b}\Big)
+73\,\frac{K-\Lambda_0 b^{2}+\dot b^{2}+2\,b\,\ddot b}{b^{2}}\nonumber\\
&\quad+5832\,d_8\,\frac{H_a^{2}}{b^{2}}\Big[\big(K+3\dot b^{2}\big)\big(\dot H_a+H_a^{2}\big)
+2H_a\,\dot b\,\ddot b\Big].
\end{align}
where we have introduced the Hubble function $H_a(t)=\dot{a}/a$. The last equation is a consequence of the first two, since due to diffeomorphism invariance the Euler-Lagrange tensor $E^{\mu}_{\ \nu}$ is identically divergenceless, and in consequence $0\equiv \nabla_\mu E^{\mu}_{\ \nu}$. 

It can be seen that imposing $b(t)=b_0$, constant, implies a first order constraint for $a(t)$ for arbitrary $p$ and $d=3$, nevertheless, we see from \eqref{eqcrucial} that when the size of the extra dimensions becomes dynamical, the number of these extra dimensions must be fixed to $p=3$ if one requires a second order system for the evolution. In consequence, requiring a ghost-free, sensible evolution for the scale factor and the size of the extra dimensions in the context of this theory, uniquely fixes the number of compact dimensions, as well as the dimension of the whole spacetime. In what follows we present different features of this theory.

\section{The solution with $b(t)=b_0$ and its perturbative stability}\label{sec:stab}
The system \eqref{constraint}, \eqref{del_a}, \eqref{del_b}, has a simple solution with a compact direction with fixed, constant radius, describing the product of de Sitter space, namely with $H(t)=H_0$ and a constant curvature space of radius $b_0$ and curvature $K=\pm1,0$. The integration constants are consistently restricted by the algebraic equations
\begin{align}\label{background}
0=&17496d_8H_0^6-5832d_8\Lambda_0H_0^4-1095H_0^2+146\Lambda_0\ ,\\ 
b_0^2=&\frac{3K}{\Lambda_0-3H_0^2}\ .
\end{align}
The second equation implies that $K$ cannot vanish, while for $K$ positive (negative), $\Lambda_0>3H_0^2$ ($\Lambda_0<3H_0^2$). The first equation of this system is a cubic equation for $H_0^2$, which for $K=+1$ has real solutions provided $\Lambda_0>0$ for arbitrary $d_8$, while for $K=-1$ it has real solutions provided $d_8>0$ with $\Lambda_0$ arbitrary in this case.

\begin{figure}[htbp]
  \centering
  \begin{minipage}{0.48\textwidth}
    \centering
    \includegraphics[width=\textwidth]{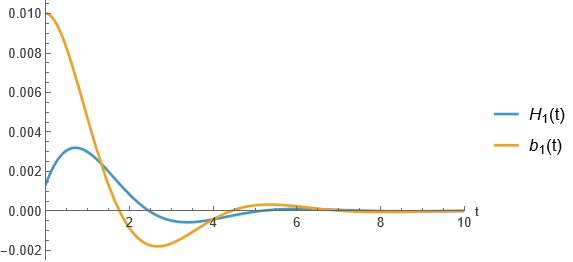}
  \end{minipage}
  \hfill
  \begin{minipage}{0.48\textwidth}
    \centering
    \includegraphics[width=\textwidth]{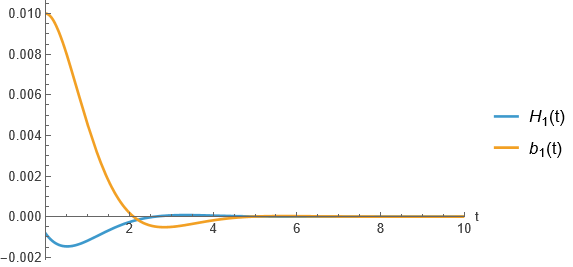}
  \end{minipage}
  \caption{\small{Evolution of the perturbation functions $H_1(t)$ and $b_1(t)$ for $K=1$, $\Lambda_0=1$, $d_8=-0.62$ (left) and $K=-1$, $\Lambda_0=1$, $d_8=0.62$ (right). For these values of the higher-curvature coupling, the background solution is characterized by the Hubble constant $H_0$ of the four-dimensional de Sitter space and a stabilized radius $b_0$ of the three-dimensional compact internal manifold $\Sigma_K$. The constraint was used to validate the results and is satisfied with residuals below $10^{-15}$.}}
  \label{fig:stab}
\end{figure}

Given the $\text{dS}_4\times\Sigma_K$ solution, with extra dimensions of a constant size given by $b_0$, we are now interested in establishing the stability of such solutions. We will start by considering perturbations that maintain the isometries of the ansatz, namely, we will evaluate the system  \eqref{constraint}, \eqref{del_a}, \eqref{del_b} in
\begin{equation}
H(t)=H_0+\epsilon H_1(t) \text{ and } b(t)=b_0+\epsilon b_1(t)\ ,
\end{equation}
keeping only first order terms in the perturbation parameter $\epsilon$. We can convert \eqref{constraint} and \eqref{del_a}, into a first-order system by introducing the redefinition $\dot{b}_1=y_1(t)$, which leads to the three-dimensional, first-order system of homogeneous ODEs, that can be written as
\[
\frac{d}{dt}\left(
\begin{array}
[c]{c}%
H_{1}\\
b_{1}\\
y_{1}%
\end{array}
\right)  =\mathbb{M}\left(
\begin{array}
[c]{c}%
H_{1}\\
b_{1}\\
y_{1}%
\end{array}
\right)\ ,
\]
with $\mathbb{M}$ a $3\times 3$ matrix given by
{\tiny
\begin{equation*}
\mathbb{M}=\left[
\begin{array}{ccc}
 H_0 \left(-\frac{21316 b_0^4}{\left(73 b_0^2+\tilde{d_8} H_0^2 K\right) \left(365 b_0^2+\tilde{d_8} H_0^2 K\right)}-4\right) & \frac{2 K \left(219 b_0^2 \left(\tilde{d_8} H_0^4-73\right)+\tilde{d_8} H_0^2 K \left(\tilde{d_8} H_0^4+73\right)\right)}{b_0 \left(73 b_0^2+\tilde{d_8} H_0^2 K\right) \left(365 b_0^2+\tilde{d_8} H_0^2 K\right)} & 73 b_0 H_0 \left(-\frac{1}{365 b_0^2+\tilde{d_8} H_0^2 K}-\frac{1}{73 b_0^2+\tilde{d_8} H_0^2 K}\right) \\
 0 & 0 & 1 \\
 73 b_0^3 H_0 \left(\frac{1}{365 b_0^2+\tilde{d_8} H_0^2 K}+\frac{1}{73 b_0^2+\tilde{d_8} H_0^2 K}\right) & \frac{146 K \left(b_0^2 \left(511-2 \tilde{d_8} H_0^4\right)+3 \tilde{d_8} H_0^2 K\right)}{\left(73 b_0^2+\tilde{d_8} H_0^2 K\right) \left(365 b_0^2+\tilde{d_8} H_0^2 K\right)} & H_0 \left(\frac{21316 b_0^4}{\left(73 b_0^2+\tilde{d_8} H_0^2 K\right) \left(365 b_0^2+\tilde{d_8} H_0^2 K\right)}-2\right) \\
\end{array}
\right]
\end{equation*}
} where for simplicity we have rescaled the higher curvature coupling such that $\tilde{d_8}=2^33^6d_8$.
The perturbative stability of the solution is ensured provided the real part of the eigenvalues of $\mathbb{M}$ is negative\footnote{See \cite{Chirkov:2020bnd} for a similar analysis in the context of Einstein-Gauss-Bonnet theory.}. Writing the characteristic polynomial and studying the Routh-Hurwitz criterion leads to the following restrictions on the dimensionless combination $d_8\Lambda_0^2$ that ensures stability
\begin{align}
K=&1 \text{ implies that } d_8\Lambda_0^2<-\frac{73}{162}\ ,\\
K=&-1 \text{ implies that } d_8\Lambda_0^2 > 0\ .
\end{align}
Figure~\ref{fig:stab} contains the evolution for $H_1(t)$ and $b_1(t)$ for different values of the coupling $d_8$ consistently leading to a stable behavior of the background solution.

\section{Non-homogeneous scalar cosmological perturbations}\label{sec:scalar}
In this section we consider cosmological scalar perturbations with a simple non-homogeneous dependence around the dS$_4\times \Sigma_K$ solution. Considering the ansatz
\begin{equation}
ds^2=-\left(1+2\Phi(t,x)\right)dt^2+\left(e^{2H_0t}-2\Psi(t,x)\right)\left(dx^2+dy^2+dz^2\right)+\left(b_0^2+2\Xi(t,x)\right)d\Sigma_K^2\ ,
\end{equation}
we define $\chi(t,x) \equiv e^{-2H_0 t}\,\Psi(t,x)$ and introduce, for simplicity, the combinations
\begin{equation}
  \Sigma \equiv H_0^2 K \tilde d_8 + 219\, b_0 , \qquad
  \Gamma \equiv \frac{16}{81}\, b_0 \tilde d_8 \left(H_0^2 b_0 + K\right).
\end{equation}
The perturbation equations coming from the components $E^t_t, E^i_j, E^a_b,E^t_{\ i}$ are respectively given by
\begin{align}
0 &= -\Gamma\, e^{-4H_0 t}\,\partial_x^4(\Phi+\chi)
   - \Sigma\, e^{-2H_0 t}\,\partial_x^2 \Xi
   + 3H_0 \Sigma\, \partial_t \Xi
   + 292\, b_0^2 e^{-2H_0 t}\,\partial_x^2 \chi \nonumber\\
  &\quad - 876\, H_0 b_0^2\,\partial_t \chi
   - 876\, H_0^2 b_0^2\, \Phi - 438\, K\, \Xi , \label{eq:tt}\\[4pt]
0 &= \Gamma\, e^{-2H_0 t}\left(\partial_t^2 + H_0 \partial_t\right)\partial_x^2(\Phi+\chi)
   - 292\, b_0^2\left[\partial_t^2 \chi + 3H_0\,\partial_t \chi + H_0\,\partial_t \Phi\right] \nonumber\\
  &\quad + \Sigma\left(\partial_t^2 + 2H_0 \partial_t\right)\Xi
   - 876\, H_0^2 b_0^2\, \Phi - 438\, K\, \Xi , \label{eq:rr}\\[4pt]
0 &= -\tfrac{1}{3}\, b_0 \Sigma\, e^{-2H_0 t}\,\partial_x^2\left(\Phi - 2\chi\right)
   - H_0 b_0 \Sigma\,\partial_t \Phi - 4H_0^2 b_0 \Sigma\, \Phi
   - b_0 \Sigma\left(\partial_t^2 + 4H_0 \partial_t\right)\chi \nonumber\\
  &\quad + 73\, b_0\left[\partial_t^2 + 3H_0 \partial_t - e^{-2H_0 t}\partial_x^2\right]\Xi
   - K\left(H_0^4 \tilde d_8 + 73\right)\Xi , \label{eq:xx}\\[4pt]
0 &= -\Gamma\, e^{-2H_0 t}\,\partial_t \partial_x^3(\Phi+\chi)
   + 292\, b_0^2\left(H_0\,\partial_x \Phi + \partial_t \partial_x \chi\right)
   - \Sigma\left(\partial_t - H_0\right)\partial_x \Xi . \label{eq:tr}
\end{align}
Notice that Equation~\eqref{eq:tr} is a total $x$-derivative. Remarkably, as anticipated, the equations for this type of scalar perturbations, that break the homogeneity of the background, contain at most second derivatives in time. This is similar to what occurs in the context of Geometric Inflation where the scalar perturbations of the four-dimensional, isotropic and homogeneous cosmologies remain of second order, in spite of containing higher derivatives with respect to the spatial directions \cite{Cisterna:2018tgx}. It would be interesting to explore cosmological perturbations of vector and tensor type since in the context of Geometric Inflation such perturbations are known to lead to instabilities \cite{Pookkillath:2020iqq}.

\section{Non-linear evolution and the time of singularity formation as a function of the coupling}\label{sec:nonlin}
In this section we consider the non-linear evolution for the ansatz
\begin{equation}
ds^2=-dt^2+a(t)^2\left(dx^2+dy^2+dz^2\right)+b(t)^2d\Sigma_K^2\ .
\end{equation}
Since the theory defined by the Lagrangian \eqref{lag} yields a constrained, second-order system for $a(t)$ and $b(t)$, it is enough to provide the initial data $(a(0),b(0),\dot{b}(0))$, because $\dot{a}(0)$ can be obtained from the constraint. In fact, the constraint is cubic in $H(t)$, so there can be more than one value of $\dot{a}(0)$ compatible with the remaining initial data.

For $K\neq0$, which is the case of interest here, this ansatz leads to singular behavior even in pure General Relativity: at a finite time $t=t_s$ the function $b(t)$ vanishes and the Kretschmann scalar diverges. That is, a singularity forms in finite time even for vanishing higher-curvature coupling, $d_8=0$. The natural question is whether this feature persists once the higher-curvature correction is included and, if so, how $t_s$ depends on the coupling $d_8$. A thorough analysis of the space of initial conditions is left for future work. Nevertheless, below we present evidence that, for positive $\Lambda_0$ and a negatively curved internal compact space, increasing the higher-curvature coupling pushes the formation of the singularity into the future.

\begin{figure}[htbp]
  \centering
  \begin{minipage}{0.48\textwidth}
    \centering
    \includegraphics[width=\textwidth]{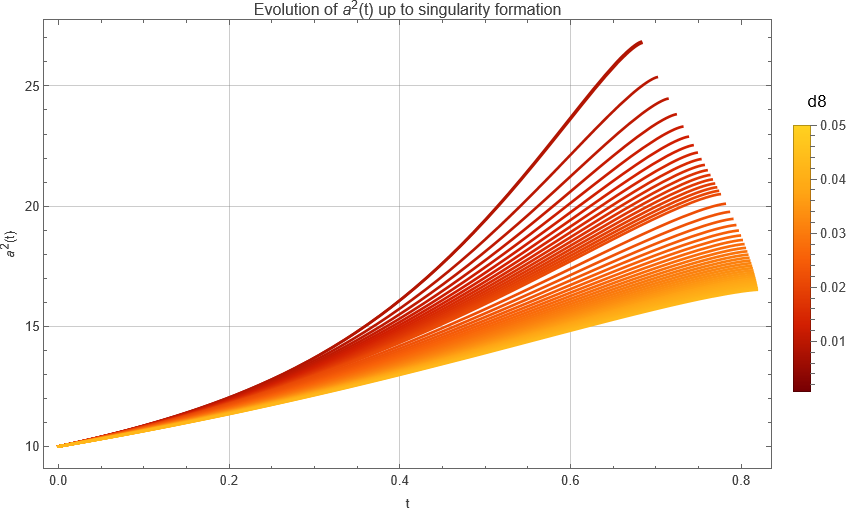}
  \end{minipage}
  \hfill
  \begin{minipage}{0.48\textwidth}
    \centering
    \includegraphics[width=\textwidth]{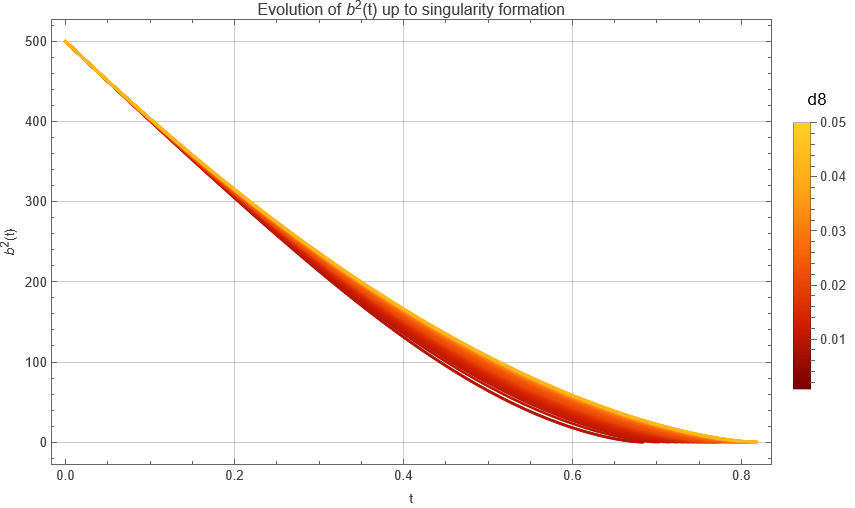}
  \end{minipage}
  \caption{\small{Evolution of the functions $a^2(t)$ and $b^2(t)$, for different values of $d_8$ until singularity formation that occurs when $b(t)\rightarrow0$. Increasing the higher curvature coupling delays the formation of the singularity. Here $\Lambda_0=0.01$ and $K=-1$.}}
  \label{fig:ab}
\end{figure}

\begin{figure}[htbp]
  \centering
  \begin{minipage}{0.48\textwidth}
    \centering
    \includegraphics[width=\textwidth]{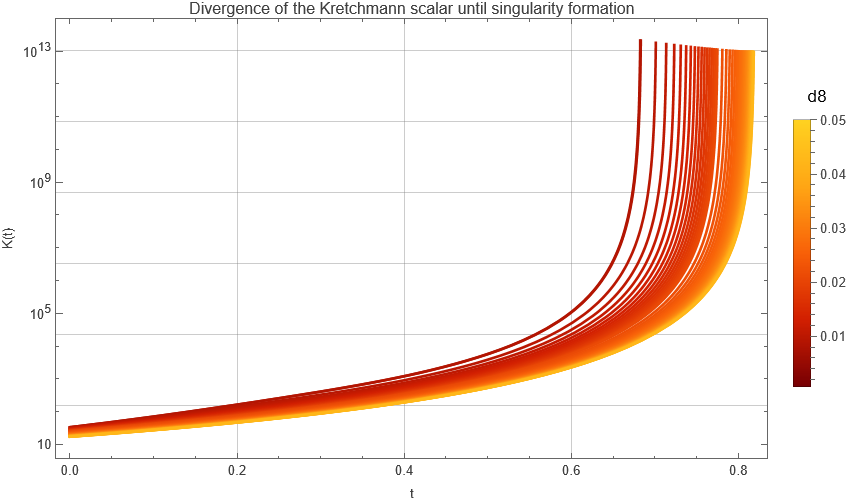}
  \end{minipage}
  \hfill
  \begin{minipage}{0.48\textwidth}
    \centering
    \includegraphics[width=\textwidth]{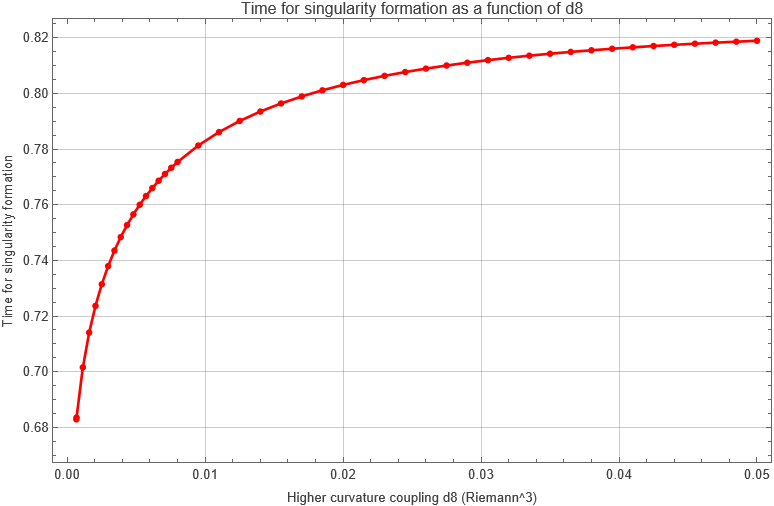}
  \end{minipage}
  \caption{\small{Evolution of the Kretschmann scalar until singularity formation (left), and dependence of the time at which the singularity is formed on the higher-curvature coupling $d_8$ (right). Here $\Lambda_0=0.01$ and $K=-1$.}}
  \label{fig:kretsch}
\end{figure}

\section{Conclusions}\label{sec:concl}
In this work we have introduced a new combination of cubic invariants in the Riemann tensor that yields second-order field equations on an FLRW ansatz times a compact space of constant curvature with radius $b(t)$. Remarkably, requiring the equations for $a(t)$ and $b(t)$ to be of second order singles out a seven-dimensional spacetime, corresponding to a four-dimensional cosmology with a three-dimensional internal space. We have shown that the solution with static $b(t)=b_0$ is linearly stable under homogeneous perturbations in an open region of parameter space. For a family of scalar cosmological perturbations that break isotropy and homogeneity, the equations also contain at most second time derivatives.

We have also studied the nonlinear evolution of $a(t)$ and $b(t)$, which develops a singularity at finite time $t=t_s$, even in General Relativity. The value of $t_s$ depends on the higher-curvature coupling, and we observe that including the cubic corrections can push the singularity a finite amount into the future. It is natural to ask whether this effect persists when corrections of order $\mathcal{R}^n$ with $n>3$ are included, and whether $t_s$ can be pushed all the way to infinity, in a manner similar to the regularization of black hole cores when an infinite tower of quasitopological corrections is included \cite{Bueno:2024dgm}. As a first step, one would need to construct a quartic combination with quasitopological-like properties on the compactified cosmological ansatz.

\acknowledgments
The authors thank Nicolás Grandi and Javier Moreno for insightful comments on related topics.  A.G. has been supported by FONDECYT Grant No. 1240247, J.O. is supported in part by the FONDECYT grant 1262452 and F. C. has been supported by FONDECYT Grant No. 1240048 and by Grant ANID EXPLORACION No. 13250014

\end{document}